\documentclass[]{spie}  

\usepackage{graphicx,amsmath,amssymb,amsfonts,array}
\usepackage[colorlinks=true, allcolors=blue]{hyperref}
\usepackage{tikz}
\usepackage{float}
\usepackage{caption}
\usepackage{color}

\newcommand{\procspie}{Proc. SPIE } 
 
\newcommand{\aap}{Astron. \& Astrophys. } 
\newcommand{\mnras}{M.N.R.A.S } 
\newcommand{\josaa}{JOSA A }

\newcommand{\cana}{\textsc{Canary} }
\newcommand{\sigdeux}[1]{\sigma^{2}_{\text{#1}}}
\newcommand{\module}[1]{\left\vert #1 \right\vert}
\newcommand{\para}[1]{\left(#1\right)}
\newcommand{\cro}[1]{\left[#1\right]}
\newcommand{\aver}[1]{\left\langle #1 \right\rangle}
\newcommand{\xth}[1]{#1^{\text{th}}}
\newcommand{\ns}{\text{n}_{\text{s}}}

\newcommand{\nf}{\text{n}_{\text{f}}}
\newcommand{\nl}{\text{n}_{\text{l}}}

\newcommand{\nngs}{\text{n}_{\text{ngs}}}
\newcommand{\hl}{h_{l}}
\newcommand{\rz}{r_0}
\newcommand{\lz}{L_0}
\newcommand{\kk}{\mathbf{k}}
\newcommand{\cnh}{C_n^2(h)}
\newcommand{\rhob}{\boldsymbol{\rho}}

\newcommand{\s}{\mathbf{S}}
\newcommand{\R}{\mathbf{R}}

\newcommand{\dipl}{\boldsymbol{\delta}_{ipl}}
\newcommand{\xipl}{x_{ipl}}
\newcommand{\yipl}{y_{ipl}}
\newcommand{\ri}{\mathbf{r}_{i}}
\newcommand{\alphap}{\boldsymbol{\alpha}_{p}}
\newcommand{\hlgsp}{h^{\text{LGS}}_{p}}
\newcommand{\djql}{\boldsymbol{\delta}_{jql}}
\newcommand{\xjql}{x_{jql}}
\newcommand{\yjql}{y_{jql}}
\newcommand{\Dijpql}{\boldsymbol{\Delta}_{ijpql}}
\newcommand{\gxx}{G_{xx}}
\newcommand{\gyy}{G_{yy}}
\newcommand{\gxy}{G_{xy}}
\newcommand{\Dphi}[1]{D_{\phi}\left(#1\right)}
\newcolumntype{P}[1]{>{\centering\arraybackslash}p{#1}}

\title{William Herschel Telescope site characterization using the MOAO pathfinder CANARY on-sky data}

\author[a]{O.A. Martin}
\author[a]{C. M. Correia}
\author[b]{E. Gendron}
\author[b]{G. Rousset}
\author[b]{F. Vidal}
\author[c]{T.J. Morris}
\author[c]{A.G. Basden}
\author[c]{R.M. Myers}
\author[a]{Y.H. Ono}
\author[a]{B. Neichel}
\author[a,d]{T. Fusco}

\affil[a]{Aix Marseille Universit\'e, CNRS, LAM, Laboratoire d'Astrophysique de Marseille, Marseille, France , 38 rue F. Joliot-Curie, 13388 Marseille Cedex 13, France}
\affil[b]{LESIA, Observatoire de Paris -- Paris Sciences et Lettres -- CNRS -- Universit\'e Paris Diderot -- Sorbonne Paris Cit\'e -- Universit\'e P. et M. Curie -- Sorbonne Universit\'e, 5 Place Jules Janssen, 92190 Meudon, France}
\affil[c]{Centre for Advanced Instrumentation, Durham Univ., South Road, Durham, DH1 3LE, UK}
\affil[d]{ONERA (Office National d'Etudes et de Recherches A\'erospatiales), B.P.72, F-92322 Ch\^atillon, France}

\authorinfo{Further author information: olivier.martin@lam.fr}

\begin{document} 
\maketitle

\begin{abstract}
\cana is the Multi-Object Adaptive Optics (MOAO) pathfinder for the future MOAO-assisted Integral-Field Units~(IFU) proposed for Extremely Large Telescopes~(ELT). The MOAO concept relies on tomographically reconstructing the turbulence using multiple measurements along different lines of sight.

Tomography requires the knowledge of the statistical turbulence parameters, commonly recovered from the system telemetry using a dedicated profiling technique. For  demonstration purposes with the MOAO pathfinder \cana, this identification is performed thanks to the Learn~\&~Apply~(L\&A) algorithm, that consists in model-fitting the covariance matrix of WFS measurements dependant on relevant parameters: $\cnh$ profile, outer scale profile and system mis-registration.

We explore an upgrade of this algorithm, the Learn 3 Steps~(L3S) approach, that allows one to dissociate the identification of the altitude layers from the ground in order to mitigate the lack of convergence of the required empirical covariance matrices therefore reducing the required length of data time-series for reaching a given accuracy. For nominal observation conditions, the L3S can reach the same level of tomographic error in using five times less data frames than the L\&A approach. 

The L3S technique has been applied over a large amount of \cana data to characterize the turbulence above the William Herschel Telescope~(WHT). These data have been acquired the 13th, 15th, 16th, 17th and 18th September 2013 and we find 0.67"/8.9m/3.07m.s$^{-1}$ of total seeing/outer scale/wind-speed, with 0.552"/9.2m/2.89m.s$^{-1}$ below 1.5~km and 0.263"/10.3m/5.22m.s$^{-1}$ between 1.5 and 20 km. We have also determined the high altitude layers above 20~km, missed by the tomographic reconstruction on \cana, have a median seeing of 0.187" and have occurred 16\% of observation time.
 
\end{abstract}


\section{INTRODUCTION}
\label{S:intro}

For Wide-Field Adaptive Optics~(WFAO) systems, the turbulence is compensated in a large Field Of View~(FoV) using a tomographic reconstruction of the phase in the entire volume~(\cite{Ragazzoni1999,Neichel2008}) or in discrete directions~(\cite{Vidal2010}), from Wave-Front Sensor~(WFS) measurements distributed over the system FoV. 

Knowledge of the statistical turbulence parameters is paramount, with several methods being dedicated to estimating them as \textsc{Slodar}~(\cite{Wilson2002,Cortes2012}), \textsc{Mass}~(\cite{Lombardi2016}) and \textsc{Scidar}~(\cite{Shepherd2014}) techniques.

As a MOAO on-sky demonstrator, \cana has been observing at the William Herschel Telescope~(WHT) since 2010~(\cite{Gendron2011}). Since an open-loop configuration is adopted, the lack of feedback between the Deformable Mirror~(DM) and WFS makes the control more sensitive to inaccuracies on the command matrix calibration. The recursive nature of a closed-loop tends to wash out these effects, which clearly is not the case here.

Vidal et al has proposed in~\cite{Vidal2010} the the Learn \& Apply~(L\&A) technique that combines both turbulence profiling and system mis-registration identification/calibration. The efficiency of this method has been demonstrated on-sky with \cana~(\cite{Martin2016WFE,Gendron2011,Vidal2014,Morris2014}). The L\&A consists in model-fitting the spatial covariance matrix of WFS measurements. The fitting is ensured thanks to the  Levenberg-Marquardt iterative algorithm which estimates a number of model-parameters, as the $\cnh$ profile, the outer scale profile and the stars position. In view of calibration, the primary model proposed by Vidal \cite{Vidal2010} included system mis-registration: pupil shifts, rotation and magnification between WFS. 

We used in the past a model \cite{Vidal2010} based on a Fourier transformation of the Von-K\'arm\'an spectrum to get the bi-dimensional (2D) spatial covariance of the wave-front slopes. The algorithm was particularly efficient when coupled to wave-front sensors with fixed sub-aperture sizes, because the computation using fast Fourier transforms (FFT) could be used to extract in one shot from the same 2D map all the covariance values between two wave-front sensors with arbitrary separation, without requiring any re-interpolation of the map -provided the sub-aperture size is a multiple of the FFT sampling. 

Now, the cone effect inherent to LGS induces layer-projected sub-aperture sizes and pupil sizes to vary with the altitude, which breaks the efficiency of the implementation when it's a matter of computing covariances between LGS and NGS. At least, an interpolation procedure could be implemented in order to access values that lay between covariance samples of the FFT, but unfortunately this kind of approximation does not go well with a fitting procedure that internally requires to compute derivatives as finite differences of the covariance function. We propose thus in Sect.~\ref{S1} a fast-computable model that includes both turbulence characteristics and system mis-registration. 

To speed up the fitting algorithm, we propose in Sect.\ref{S:l3s} an upgrade to the L\&A, called the \emph{Learn 3 steps~(L3S)}, that splits the identification of the ground layer parameters from those of altitude layers. The ground layer is known to  dominate the turbulence strength with relatively slow time-variations~(\cite{Martin2014}). It appears thus that a large time-span of accumulated data would be required to obtain statistical convergence on the generation of the covariance matrices of phase~(\cite{Martin2012}), the latter being an input to the tomography. In splitting the ground and altitude parts, we mitigate the  mis-convergence of statistics obtained from system telemetry. 

In addition, we have included in the L3S a separate estimation of the wind-speed profile. From the turbulence profile, we reconstruct the phase at each retrieved altitude using an MMSE approach. The wind-speeds on those layers is then determined from the Full Width at Half Maximum~(FWHM) of the average temporal auto-correlation function of slopes. 

We present then in Sect.~\ref{S:stats} a characterization of the turbulence along altitude above the WHT from the \cana on-sky data acquired in September 2013. We have processed about 2,000 data sets with the L3S algorithm to present plots of distribution of the seeing, outer scale and wind-speed. We have also characterized the median values for the ground layer~(below 1.5\,km) the altitude layers~(between 1.5 and 20~km) and the very high altitude layers~(above 20~km) inaccessible to tomography with \textsc{Canary}. We conclude in Sect.~\ref{S:conclusions}.

\section{Modeling the spatial covariance of WFS slopes on top of mis-registration}
\label{S1}
\subsection{Analytic expressions of the covariance} 
\label{SS:model}

\begin{minipage}{.45\columnwidth}
	\includegraphics[scale=.3]{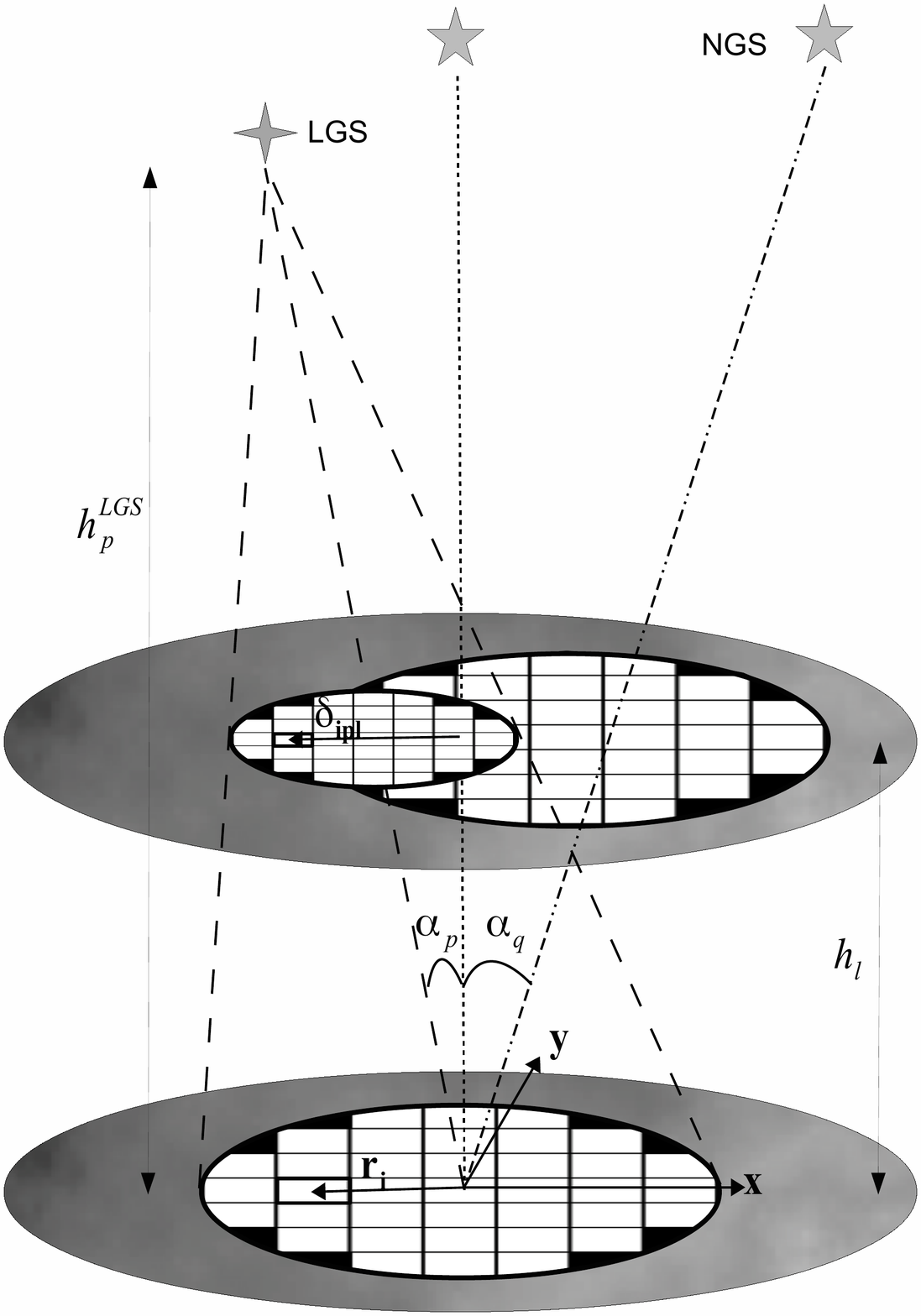}
	\label{F1}
\end{minipage}
\begin{minipage}{.45\columnwidth}
	\centering
	\begin{tabular}{lP{6.5cm}}
		\hline
		Variable & Description\\ 
		\hline
		$\ri$ & Coordinates of the sub-aperture $i$ in the pupil plane\\ 
		$\alphap$ & Angular position of $\xth{p}$ stars from the center of the FoV\\
		$\hl$ & altitude of the $l^{\text{th}}$ turbulent layer\\
		$\dipl$ & Coordinates of the $\xth{i}$ sub-aperture of the $\xth{p}$ WFS at the altitude $h_l$\\
		$\xipl$ & Component of $\dipl$ over the x-axis\\
		$\yipl$ & Component of $\dipl$ over the y-axis\\
		$\hlgsp$& Focus altitude of the $\xth{p}$ WFS. \\
		$d_i$& Size of the $\xth{i}$ sub-aperture in the pupil plane \\
		$\Delta x_p,\Delta y_p$ & Pupil shifts for the $\xth{p}$ WFS\\ 
		$\theta_p$ & Rotation shift for the $\xth{p}$ WFS\\ 
		$M_p$ & Magnification for the $\xth{p}$ WFS\\ 
		\hline
	\end{tabular}
	\label{T:2}
\end{minipage}
\captionof{figure}{\small{\textbf{Left:} Illustration of the geometry of the tomographic problem. For LGS WFSs, the pupil footprint decreases with altitude because of the cone effect. The coordinates of each LGS slope are affected by a homothety factor. \textbf{Right:} Notations used along this paper.}}
\vspace{.25cm}

In this proceeding, we review the calculations from \cite{Martin2012,Gendron2014} in a more general way, in including system mis-registration.

Consider the $\xth{i}$ sub-aperture of the $\xth{p}$ WFS. In the pupil plane, we note $\ri$ the position of the $\xth{i}$ sub-aperture which does not depend on the actual WFS since all are conjugated at the pupil plane. We include DM/WFS mis-registrations: for the $\xth{p}$ WFS, we define $\Delta x_{p}$ and $\Delta y_{p}$ as the x and y pupil shifts, $\theta_p$ as the rotation and $M_p$ as the magnification. We note $\ri^\prime$ as the new $\xth{i}$ sub-aperture's coordinates in the pupil:
\begin{equation}
	\ri^\prime = \left[
	\begin{aligned}
		& \cos(\theta_p) \quad -\sin(\theta_p)\\
		& \cos(\theta_p) \quad +\sin(\theta_p)\\
	\end{aligned}
	\right]
	\times \cro{M_p\times \ri} + \Delta x_p + \Delta y_p
\end{equation}

We note $\alphap$ as the angular position of the $\xth{p}$ WFS from the center of the FoV. We will finally note $\hlgsp$ the altitude focus of guide stars and $\dipl$ as the coordinate of this sub-aperture, at the $\xth{l}$ layer, in the pupil plane. We have~:
\begin{equation} \label{E:3}
	\dipl= \left\lbrace
	\begin{aligned}
		& \alphap \hl + \ri^\prime \para{1 - \frac{ \hl }{\hlgsp}} \text{ for a LGS WFS and $\hl \leq \hlgsp$}.\\
		& \alphap \hl + \ri^\prime \text{ for a NGS WFS}
	\end{aligned}
	\right.
\end{equation}
For the following theoretical developments, we define $\xipl$ and $\yipl$ the projection of $\dipl$ on the reference pupil plane:
\begin{equation} \label{E:4}
	\dipl = \xipl \mathbf{x} + \yipl \mathbf{y},
\end{equation}
where $\mathbf{x}$ and $\mathbf{y}$ are forming an orthonormal basis along directions $x$ and $y$.
For a Shack-Hartmann WFS, the 2D slopes' map is obtained from the average phase gradient over the lenslet~:
\begin{equation} \label{E:S_dipl}
	\mathbf{S}(\dipl) = \dfrac{1}{d_i}\iint_{\mathcal{A}} \boldsymbol{\nabla}\boldsymbol{\phi}(u_i,v_i) du_idv_i,
\end{equation}
where $d_i$ is the $\xth{i}$ sub-aperture size, i.e. the edge to edge distance in the pupil plane considering all sub-apertures are squared, and $\mathcal{A}$ is the square integration domain comprised between $-d_i/2$ and $d_i/2$ in $x$ and $y$ directions. The integration variables $u_i$ $v_i$ are related to the $\xth{i}$ sub-aperture and are given by $u_i = x - \xipl$ and $v_i = y-\yipl$. 

The x-axis and y-axis slopes are the scalar product between $\mathbf{S}(\dipl)$ and respectively $\mathbf{x}$ and $\mathbf{y}$. In projecting the slopes along x or y directions, the integral on sub-aperture surface in Eq.~\ref{E:S_dipl} becomes integrals along sub-aperture side~:
\begin{equation} \label{E:S_dipl2}
	\begin{aligned}
		& \mathbf{S}(\dipl).\mathbf{x} = \dfrac{1}{d_i}\int\limits_{-d_i/2}^{d_i/2}  dv_i \para{\phi(\xipl+d_i/2,v_i) - \phi(\xipl-d_i/2,v_i)}\\
		&\mathbf{S}(\dipl).\mathbf{y} = \dfrac{1}{d_i}\int\limits_{-d_i/2}^{d_i/2}  du_i \para{\phi(u_i,\yipl+d_i/2) - \phi(u_i,\yipl-d_i/2)}.
	\end{aligned}
\end{equation}
We then define the spatial covariance $\gxx$, $\gyy$ and $\gxy$ as following:
\begin{equation}
	\begin{aligned}
		&\gxx(\Dijpql) = \aver{(\mathbf{S}(\dipl).\mathbf{x}) (\mathbf{S}(\djql).\mathbf{x})^t  }\\
		&\gyy(\Dijpql) = \aver{(\mathbf{S}(\dipl).\mathbf{y}) (\mathbf{S}(\djql).\mathbf{y})^t  }\\
		&\gxy(\Dijpql) = \aver{(\mathbf{S}(\dipl).\mathbf{x}) (\mathbf{S}(\djql).\mathbf{y})^t  },
	\end{aligned}
\end{equation}
where $\Dijpql = \dipl - \djql$ is the separation between the $\xth{i}$ sub-aperture of $\xth{p}$ WFS and the $\xth{g}$ sub-aperture of $\xth{q}$ WFS, at altitude $\hl$. We develop here the method to get $\gxx(\Dijpql)$ only; both $\gyy(\Dijpql)$ and $\gxy(\Dijpql)$ are similarly derived. Noting the averaged gradient is the difference of phase along each side of the sub-aperture, we then get~:
\begin{equation} \label{E:SxSx}
	\begin{aligned}
\gxx(\Dijpql) = \dfrac{1}{d_i d_j} \iint dv_idv_j  &\left\langle \para{\phi(\xipl+d_i/2,v_i + \yipl) - \phi(\xipl-d_i/2,v_i+\yipl)}\right.\\
&\left.\para{\phi(\xjql+d_j/2,v_j+\yjql) - \phi(\xjql-d_j/2,v_j+\yjql)}\right\rangle. 
	\end{aligned}
\end{equation}
We now use the identity:
\begin{equation} \label{E:idenremar}
2\para{A -a}\para{B-b} =  -(A-B)^2 + (A-b)^2 +(a-B)^2 -(a-b)^2,
\end{equation}
which allows to rewrite Eq.~\ref{E:SxSx} as follows:
\begin{equation} \label{E:SxSx3}
\begin{aligned}
\gxx(\Dijpql)= \frac{1}{2d_id_j} \iint dv_idv_j &\left\langle  -(\phi(\xipl+d_i/2,v_i+\yipl)-\phi(\xjql+d_j/2,v_j+\yjql))^2 \right.\\
& -(\phi(\xipl-d_i/2,v_i+\yipl)-\phi(\xjql-d_j/2,v_j+\yjql))^2 \\
&+ (\phi(\xipl+d_i/2,v_i+\yipl) -\phi(\xjql-d_j/2,v_j+\yjql))^2\\
&\left. + (\phi(\xipl-d_i/2,v_i+\yjql) -\phi(\xjql+d_j/2,v_j+\yjql))^2\right\rangle\\
	\end{aligned}
\end{equation}
Considering $\Dphi{\boldsymbol{\rho}} = \aver{(\phi(\mathbf{r})-\phi(\mathbf{r} + \boldsymbol{\rho}))^2}$ as the phase Structure Function~(SF), we have the final expression of the x-axis slopes covariance~: 
\begin{equation} \label{E:Gxx}
	\begin{aligned}
&\gxx(\Dijpql) = \dfrac{1}{2d_id_j} \iint dv_idv_j \left(-2\Dphi{ \Dijpql + \dfrac{d_i-d_j}{2} \mathbf{x} + (v_i -v_j)\mathbf{y} }\right.\\
& \left. +\Dphi{\Dijpql+ \dfrac{d_i+d_j}{2}\mathbf{x}+ (v_i -v_j)\mathbf{y} } + \Dphi{ \Dijpql - \dfrac{d_i+d_j}{2}\mathbf{x} + (v_i - v_j)\mathbf{y} } \right) + t_{xx},
	\end{aligned}
\end{equation}
where $t_{xx}$ is a constant value to include the telescope tracking error and vibrations we assume to provide an additional isoplanatic tip-tilt to the turbulence. 
We now consider $W_\phi(\kk)$ as the Von-K\'arm\'an spatial Power Spectral Density~(PSD) given by~:
\begin{equation} \label{E:wphi}
	W_\phi(\kk) = 0.023 \rz^{-5/3} \para{\module{\kk}^2 + 1/\lz^2}^{-11/3},
\end{equation}
The phase Structure Function~(SF) is related to the spatial PSD $W_\phi(\kk)$ by~:
\begin{equation} \label{E:dphi2}
	\Dphi{\rhob} =  2\iint_{\mathbb{R}^2}W_\phi(\kk)\para{1 - e^{2i\pi\kk\rhob}}d\kk.
\end{equation}
where the phase SF following the von-K\'arm\'an statistics is~:
\begin{equation}\label{E:Dphi}
D_{\phi}(\boldsymbol{\rhob}) = \frac{2^{1/6} \Gamma(11/6)}{\pi^{8/3}} \cro{\frac{24}{5}\Gamma(6/5)}^{5/6} \para{\dfrac{\lz(\hl)}{\rz(\hl)}}^{5/3} \cro{\frac{\Gamma(5/6)}{2^{1/6}}-  \rhob^{5/6} K_{5/6}(\rhob)},
\end{equation}
with $\rho = 2\pi \module{\Dijpql}/\lz(\hl)$. In the same way, we get:
\begin{equation} \label{E:Gyy}
	\begin{aligned}
&\gyy(\Dijpql)=\frac{1}{2d_id_j} \iint du_idu_j \left(-2\Dphi{ \Dijpql + \dfrac{d_i-d_j}{2}\mathbf{y} + (u_i -u_j)\mathbf{x}}\right.\\
& \left.+\Dphi{\Dijpql+ \dfrac{d_i+d_j}{2}\mathbf{y}+ (u_i -u_j)\mathbf{x}} +\Dphi{ \Dijpql - \dfrac{d_i+d_j}{2}\mathbf{y} + (u_i - u_j)\mathbf{x} } \right) + t_{yy},
\end{aligned}
\end{equation}
and
\begin{equation} \label{E:Gxy} 
	\begin{aligned}
\gxy(\Dijpql)=\frac{1}{2d_id_j} \iint du_j dv_i & \left(-\Dphi{ \Dijpql - \para{u_j-\dfrac{d_j}{2}}\mathbf{x} +  \para{v_i-\dfrac{d_i}{2}}\mathbf{y} }\right.\\ 
&- \Dphi{ \Dijpql -\para{u_j+\dfrac{d_j}{2}}\mathbf{x} + \para{v_i+\dfrac{d_i}{2}}\mathbf{y} } \\
&+ \Dphi{ \Dijpql - \para{u_j-\dfrac{d_j}{2}}\mathbf{x} + \para{v_i+\dfrac{d_i}{2}}\mathbf{y}}\\
		&\left.+\Dphi{ \Dijpql -\para{u_j+\dfrac{d_j}{2}}\mathbf{x} + \para{v_i-\dfrac{d_i}{2}}\mathbf{y}} \right) + t_{xy}.
	\end{aligned}
\end{equation}

To retrieve the solution, the integrals need be discretized along sub-apertures sides. In~(\cite{Martin2012,Gendron2014}), this issue is solved using a mid-point version of the Hudgin model~: the phase gradient is calculated in taking only the mid-point edge-to-edge phase difference. In such a case, instead of having the integrand variables $u_i$ and $v_i$ ranging from $-d_i/2$ to $d_i/2$, they are evaluated at a single point, equating the integral to zero. In so doing we find again the analytic expression  in~(\cite{Martin2012,Gendron2014}). This approach allows to drastically reduce the computation time of analytic covariance matrices, few seconds with 8 7$\times$7 SH WFS on \textsc{Canary}. The cost is an overestimation of 15\% on the covariance peak~(\cite{Martin2012}), but the wings of the spatial covariance are still well reproduced. Overall the covariance model depends on the following parameters~: 
\begin{itemize}
	\item $\nl$: number of discrete layers.
	\item $\hl$: altitude layers.
	\item $r_0(\hl)$: $r_0$ value for the discrete altitude at altitude $\hl$
	\item $L_0(\hl)$: $L_0$ value for the discrete altitude at altitude $\hl$
	\item $\Delta x_p$, $\Delta y_p$, $\theta_p$ and $M_p$: system mis-registration.
	\item $t_{xx}$, $t_{yy}$ and $t_{xy}$: constant value on the covariance map for taking into account telescope tracking error and vibrations.
\end{itemize}

\subsection{Covariance of the aliasing }
\label{A:2}

The covariance of the aliased phase can be derived using Eqs.~\ref{E:Gxx},~\ref{E:Gyy} and~\ref{E:Gxy}. These equations involve an expression of the phase SF which depends on the turbulence parameters and the system geometry -- Eq.~\ref{E:Dphi}. To get the aliasing contribution in the covariance, we have to split the phase into a parallel part, i.e. compensated by the system, and an orthogonal part:
\begin{equation}
	\phi = \phi^\parallel + \phi^\perp.
\end{equation}
The orthogonal phase contains only spatial frequencies higher than $1/2d$, the DM cut-off frequency. In the case of a Fried geometry~(most frequent configuration using SH WFS), the WFS sub-aperture spacing is equal to the DM actuators pitch. Considering then $\lz >> 2d$, the aliasing SF $D^\perp_\phi(\rhob)$ becomes
\begin{equation} \label{E:dphialias}
	D^\perp_\phi(\rhob) =  0.046 \rz^{-5/3}\iint \limits_{1/2d}^{\quad\infty} \kk^{-11/3} (1 -e^{-2i\pi \kk\rhob}) d\kk.
\end{equation}
Using Eq.~\ref{E:dphialias} to replace the SF expression in Eqs.~\ref{E:Gxx},~\ref{E:Gxy} and~\ref{E:Gxy}, we get an analytic expression of the spatial covariance of the aliasing.

\section{Learn 3-steps approach}
\label{S:l3s}

\subsection{L3S concept}

For determining the model parameters listed in Sect.~\ref{SS:model}, we use an extension of the Learn\&Apply algorithm. This extension is called the \emph{Learn 3 steps} and operates similarly to the latter.\\

The main idea of the Learn technique is to least-squares fit the empirical covariance matrix from telemetry to the theoretical one. To mitigate the presence of noise we exclude the diagonal of these matrices during the minimization, for which we employ an iterative Levenberg-Marquardt Algorithm~(LMA). This method has been demonstrated to be powerful for performing the tomography on \cana~(\cite{Gendron2011}).\\

\begin{figure}[h!]
	\begin{center}
		\begin{tikzpicture}{x=1cm,y=1cm,>=stealth}
		\draw[line width = .05cm, color=blue] (0,0) rectangle (4,3.1);
		 \draw (0.1,3.1) node[below right,text width =4 cm, color=blue]{\textbf{NGS/LGS cov. mat.}:\\
			\vspace{.25cm}
			
			NGSs slopes\\
			LGSs slopes\\
			Stars positions $\boldsymbol{\alpha}_p$\\
			LGSs altitude $\hlgsp$\\
			Prior on $\nl$\\
		};
		\draw[->, color=blue] (4,2) -- (5,2);
		\draw (5,2.75) rectangle node[text width =1cm,text centered]{Rm. TT} (6,1.25);
		\draw[->, color=blue] (6,2) -- (7,2);
		\draw (7,2.75) rectangle node[text width =1cm,text centered]{Rm. mean slopes} (8,1.25);
		
		\draw[->, color=red] (4,-1) -- (4.5,-1) -- (4.5,1.5) -- (5,1.5);
		\draw[->, color=red] (6,1.5) -- (7,1.5);
		\draw[line width = .05cm, color=red] (0,-.5) rectangle (4,-3.5); 
		\draw (0.1,-.5) node[below right,text width =4cm, color=red]{\textbf{cov. mat model}:\\
			\vspace{.25cm}
			
			$\hl,r_0(\hl)$,$L_0(\hl)$\\
			$t_{xx},t_{yy}$ and $t_{xy}$\\
			$\Delta x_p,\Delta y_p,M_p,\theta_p$\\			
		};

		\draw[->, color=blue] (8,2) -- (8.5,2) -- (8.5,0.75) -- (7.25,0.75) -- (7.25,0.5);
		\draw[->, color=red] (8,1.5) -- (8.25,1.5) -- (8.25,0.85) -- (7,0.85) -- (7,0.5);
		
		\draw[->, color=blue] (5.25,1.25) -- (5.25,-.5);\draw[->, color=red] (5.5,1.25) -- (5.5,-.5);
		\draw[->, color=black!60!green] (6,0) -- (6,-.5);\draw[->, color=black!60!green] (6,-1) -- (6,-1.5);
		\draw[->, color=red] (4,-1.85) -- (5,-1.85); \draw[->, color=blue] (4,.5) -- (4.25,.5) -- (4.25,-1.65) -- (5,-1.65);
		
		\draw[line width = 0.05 cm,fill=white] (5,0) rectangle node{LMA} (7.5,.5);

		\draw[line width = 0.05 cm,fill=white] (5,-1) rectangle node{LMA} (7.5,-.5);

		\draw[line width = 0.05 cm,fill=white] (5,-2) rectangle node{LMA}(7.5,-1.5);

		\draw[dashed,] (8.75,-4) -- (8.75,4);
		\draw (10.25,3.75) node{\textbf{Outputs}};
		\draw (6.25,3.75) node{\textbf{Learn 3 steps}};
		\draw (1.75,3.75) node{\textbf{Intputs}};
		\draw[dashed] (4.25,-4) -- (4.25,4);
		
		\draw[->,color=black!60!green] (7.5,.25) -- (9,.25);\draw[dashed,color=black!60!green] (9,0) rectangle node[text width =2.5cm,color=black!60!green]{\textbf{Step 1}:\\ $\hl,r_0(h),L_0(h)$, $h>0$}(12,1.5);

		\draw[->,color=black!60!green] (7.5,-.75) -- (9,-.75);\draw[dashed] (9,-1.) rectangle node[text width=2.5cm,color=black!60!green]{\textbf{Step 2}: $r_0(0),L_0(0)$} (12,0);
		\draw[->,color=black!60!green] (7.5,-1.75) -- (9,-1.75);\draw[dashed,color=black!60!green] (9,-2.5) rectangle node[text width=2.5cm,color=black!60!green]{\textbf{Step 3}:\\
			$t_{xx},t_{yy}$ and $t_{xy}$\\
			$\Delta x_p,\Delta y_p,M_p,\theta_p$\\
		} (12,-1.);
		\draw[line width =0.05cm] (4.4,3.1) rectangle (8.6,-3.5);
	
		\end{tikzpicture}
	\end{center}
	\caption{\small{Schematic of the Learn 3 steps algorithm. In blue, red and dark green, we have the paths of respectively the covariance matrix from measurements, from model and the model parameters. On the left, the inputs are the telemetry and the phase gradient covariance matrix model. On the center, the algorithms operates several iterative LMA model-fitting of the measurements' covariance matrix. See text for further details.}}
	\label{F:L3steps}
\end{figure}
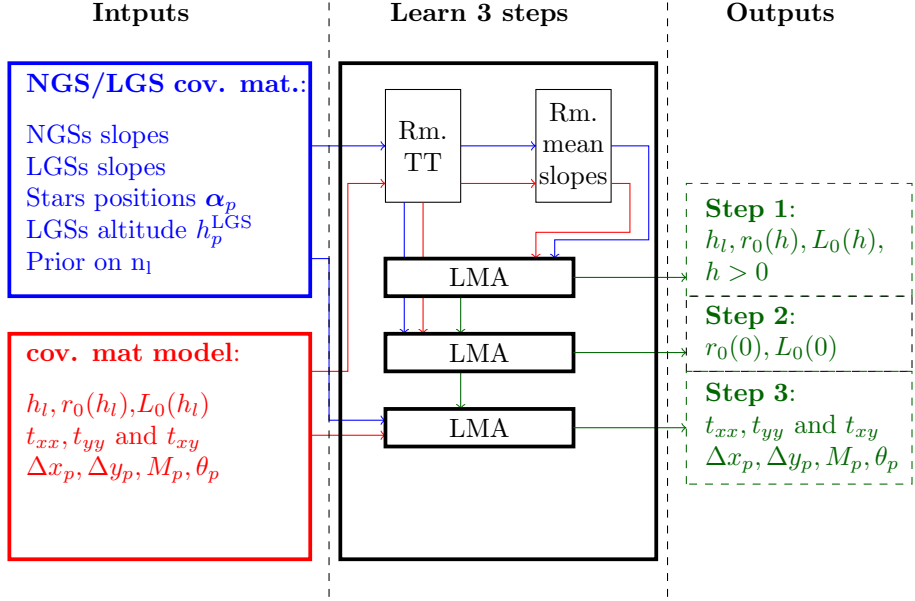

With the L3S we try to reduce the sensitivity of the model-fitting procedure to the lack of convergence we have on the empirical covariance using short time-series of data. We aim to reach the same tomographic error on a shorter time series than we would with the retrieved parameters of the original method. From the following observation~(\cite{Martin2012}): the convergence effect is stronger when the turbulence gets a larger seeing and a slower wind, we take it that the ground layer is the one which contributes the most to this effect. We propose to retrieve it independently from the altitude layers. We report in Fig.~\ref{F:L3steps} a diagram of the L3S algorithm:
\begin{itemize}
    \item \textbf{Step 1:} we start by removing the tip-tilt mode in measurement and model for each WFS. Then we remove the mean slopes between all the WFS corresponding to the ground layer~(GL) contribution in both measurement and model. In doing that we suppress the GL contribution of the covariance matrices in order to focus on the altitude layer identification. By applying the LMA on such matrices, we get an estimation of $r_0(\hl)$ and $\lz(\hl)$ in altitude.
    \item \textbf{Step 2:} We get back to the initial values of covariance matrices, always tip-tilt removed, in keeping the same parameters in altitude retrieved during the previous step. The LMA gives then an estimation of the ground layer seeing and outer scale.
    \item \textbf{Step 3:} Finally, we add back the tip-tilt components in the measurements and model covariance matrices. We then retrieve the telescope tracking error and vibrations coefficients $t_{xx},t_{yy}$ and $t_{xy}$. They simulate an isoplanatic tip-tilt added to the turbulent one. During this third step, we have also the possibility to retrieve the system mis-registration $\Delta x_p$, $\Delta y_p$, $M_p$ and $\theta_p$. 
\end{itemize}

\subsection{Performance comparison}
\label{SS33}

We compare the L3S approach with the initial L\&A approach proposed by Vidal~(\cite{Vidal2010}). For only considering the gain on spitting the identification of relevant parameters in three steps, we use exactly the same covariance matrix model in the both cases. We call thus \emph{Learn 1 step}~(L1S) the version of the L3S where we retrieve all the parameters during a single step.

We have selected three data sets acquired by \cana in 2013, comprising 10,000 frames sampled at 150~Hz. Data sets are the concatenation of slopes measured by three off-axis NGS WFS in a 2' field, one one-axis WFS and four Rayleigh LGS WFS. These later were focused at 21~km and positioned at 23'' off-axis in a square configuration~(\cite{Morris2014})\\

We report in Tab.~\ref{T:glbparam} the observation conditions we got for these three cases. To be more accurate on the estimation of the turbulence parameters during the observation, all of these values have been evaluated on the entire 10,000 frames.\\

\begin{table}[h!]
	\begin{center}
		\begin{tabular}{|c|c|c|c|}
			\hline
			Case &  1& 2 & 3 \\
			\hline
			Date & 09/16/2013 & 09/17/2013  & 09/18/2013 \\
			\hline 
			Hour & 23h58m& 00h13m &  23h16m \\
			\hline
			\hline
			Seeing (") & 1.40 & 0.89  & 0.310  \\
			Altitude seeing  & 0.20 & 0.27  & 0.12  \\
			Ground seeing & 1.37 & 0.82 &  0.26 \\
			\hline
		\end{tabular}
	\end{center}
	\caption{\small{Table of observation conditions for the three studied cases. The ground layer gathers all the layer below 1~km, that corresponds to the tomographic resolution we have with \cana~(\cite{Vidal2014}).}}
	\label{T:glbparam}
\end{table}

We have split each of these data sets into two parts: the 5,000 first frames form the \emph{calibration set} while the 5,000 last form the \emph{test set}. We select $\nf$ frames in the calibration set to compute the empirical covariance matrix. In using either the L\&A or L3S method, we retrieve two different sets of model parameters. Then, from the model-fitted covariance matrices, we derive a MMSE reconstructor $\R$ and then a Wave Front~(WF) \emph{tomographic error} in nm rms~(\cite{Gendron2014}), in applying this reconstructor on the test set. We redo these operations for different values of $\nf$ in order to get the evolution of the tomographic error as function of $\nf$. \\

\begin{figure}[h!]
	\begin{center}
		\includegraphics[scale=.42]{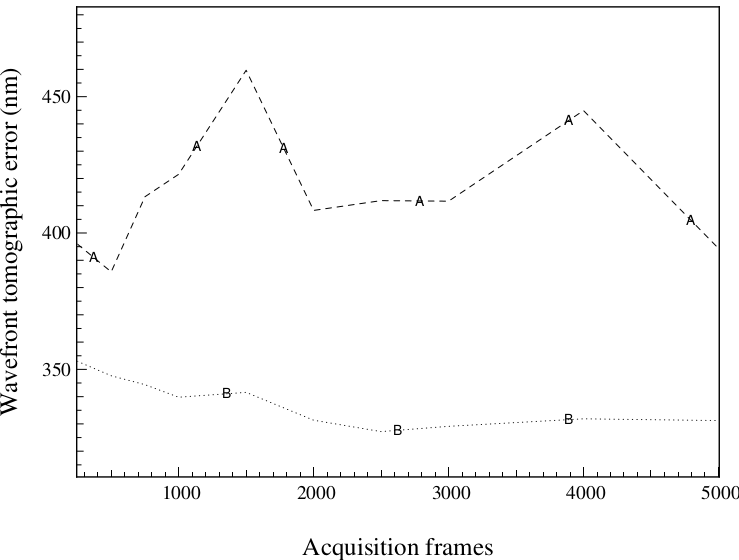}
		\hspace{.15cm}
		\includegraphics[scale=.42]{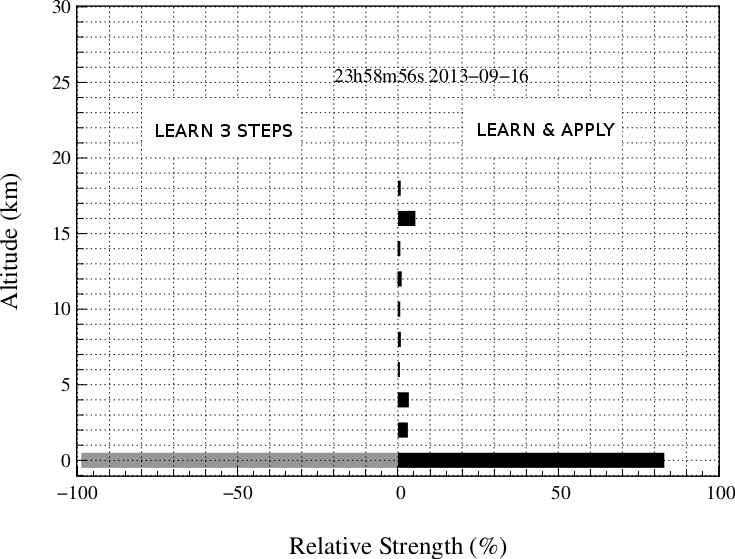}
		\vspace{.75cm}
		
		\includegraphics[scale=.42]{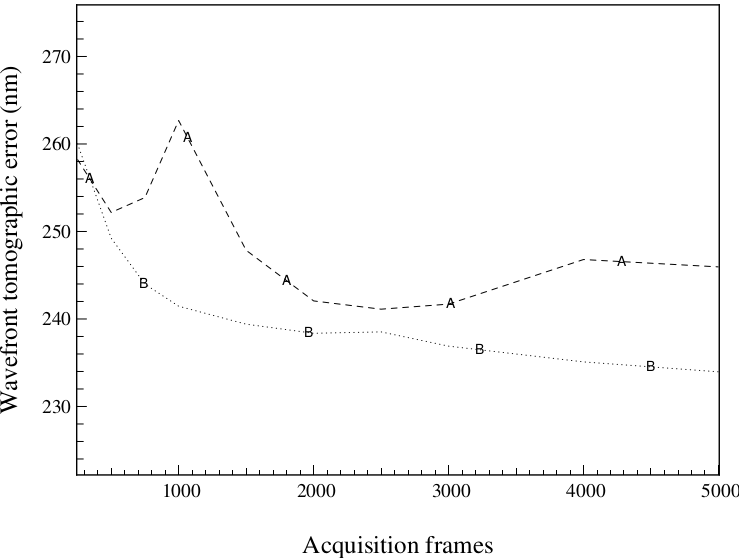}
		\hspace{.15cm}
		\includegraphics[scale=.42]{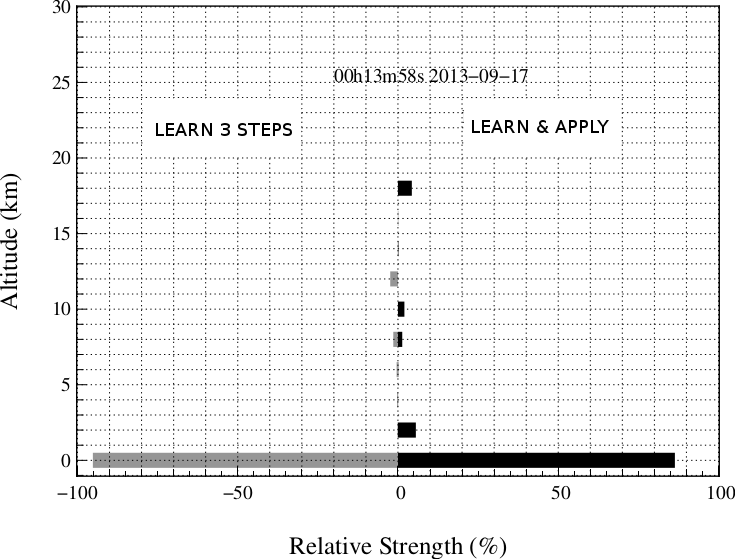}	
		\vspace{.75cm}
		
		\includegraphics[scale=.42]{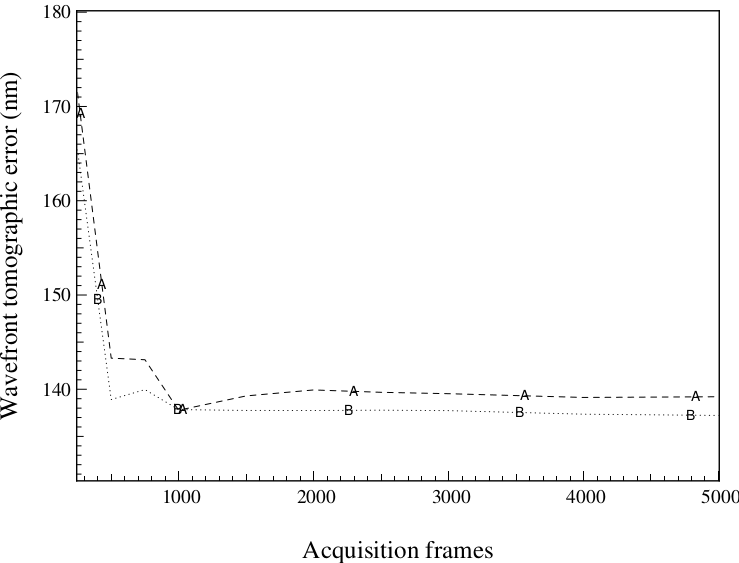}
		\hspace{.15cm}
		\includegraphics[scale=.42]{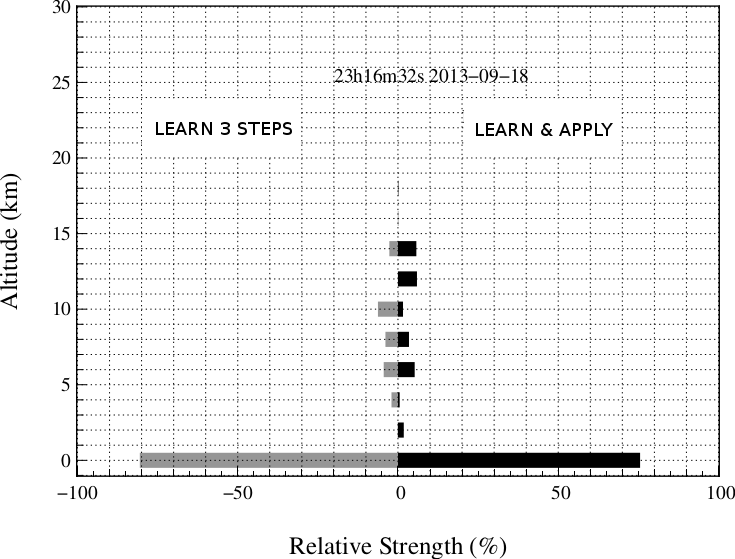}
		
	    \caption{ \small{\textbf{Left~:} Tomographic error got using both the L\&A and L\&A 3-steps algorithms for retrieving the turbulence properties, for three different cases where the seeing was respectively equal to 1.4, 0.9 and 0.3 arcsec. A: Learn-1-step approach B: Learn 3-steps. \textbf{Right~:} Comparison between profiles retrieved using either L\&A algorithm or Learn 3-steps approach. See text for more explanations.}}
	    \label{F:compar}
	\end{center}
\end{figure}

In Fig.~\ref{F:compar}, we show the tomographic error versus $\nf$ for both L1S and L3S, for the three different sets. In addition in Fig.~\ref{F:compar}, we compare turbulent profiles retrieved with $\nf=5,000$ frames for both methods. We get an improvement of the tomographic error on the three cases following two criteria:
\begin{itemize}
    \item \textbf{Better robustness to the lack of convergence of empirical covariance~:} tomographic errors with the L3S method are decreasing faster than the L1S ones. It means we reach the same level of accuracy on the turbulence parameters identification in using around five times less frames in nominal conditions.
    \item \textbf{Better accuracy~:}  As it appears in Fig.~\ref{F:compar}, tomographic error plateaus when the time-series length is large enough. The final value of tomographic error we get with the L3S is smaller than the one given by the L1S method. For a time-series of 5,000 frames, we get an improvement on the quadratic error of 50\%, 30\% and 12\% for sets 1, 2 and 3 respectively. This gain is thus increasing with the seeing. Indeed, for worse observation conditions than nominal ones, the mis-convergence of turbulence statistic parameters impacts more the L1S parameter retrieval.
    
    The same is confirmed by the turbulent profiles retrieved by the two approaches~: for data set 1, the seeing is worse than nominal conditions~(median of 0.7") but the L\&A has retrieved weak altitude layers. In comparing with the L3S results, they are fake layers coming from a lack of convergence of the LMA which did not reach the optimal solution. It would be necessary to get many more frames to make the L1S algorithm converging towards this solution, which is in that case a very strong ground layer only. The L3S allows to identify this ground layer only from few thousands frames only.
\end{itemize}

\subsection{Wind-speed profiling}
\label{S:windspeed}

The temporal properties of the turbulence are required for data analysis, LQG control~(\cite{Sivo2015}) and PSF reconstruction~(see paper 9909-64 \emph{PSF reconstruction validated using on-sky \cana
data in MOAO mode} in that conference). In that last case, considering we are seek to reconstructing a long-exposure PSF, the knowledge of wind direction is no longer required. The long-exposure PSF results from an average over all directions. We thus only need to estimate the wind-speed profile.

We discuss here another approach than ones developed from external profiler as SCIDAR~(\cite{Osborn2015}) or SLODAR~(\cite{Butterley2006}) based on a cross spatio-temporal covariance fitting. It is based on the temporal auto-covariance function of slopes acquired on NGS WFS in loop disengaged. Such a technique has been also used on \textsc{Raven}, the Subaru telescope technical and science MOAO demonstrator~(~\cite{Ono2016}). We start by reconstructing the WFS slopes layer-by-layer, using the knowledge of the vertical distribution of the turbulence we learn previously fron the L3S approach. Then, we determine the FWHM of the temporal auto-correlation function of slopes, which is related to the wind speed according a calibrated relation.

\subsubsection{Wind-speed measurements from FWHM temporal auto-correlation function}
\label{SS:FWHM}

We note $\s(i,p)$ as the $\xth{i}$ measurements provided by the $\xth{p}$ WFS. It is a $\nf$-size vector where $\nf$ is the time-series length. The temporal auto-covariance function of $\s(i,p)$  is $\boldsymbol{\gamma}(i,p)$ and comes from a two-dimensional Fourier transform:
\begin{equation} \label{E:g}
\boldsymbol{\gamma}(i,p,\Delta t) = \mathcal{F}^{-1}\para{\module{\mathcal{F}\para{\s(i,p)}}^2}(\Delta t)
\end{equation}
We then average the $\gamma(i,p,\Delta t)$ functions over $x$ and $y$-axis measurements and NGS WFS to get the $\nf$-sized vector $\overline{\gamma}$:
\begin{equation} \label{E:gavg}
\overline{\gamma}(\Delta t) = \dfrac{1}{\ns\nngs}\times\sum_{p=1}^{\nngs}\sum_{i=1}^{\ns} \para{\gamma(i,p,\Delta t) - \sigdeux{$\eta$}(i,p)},
\end{equation}
where $\gamma(i,p,\Delta t)$ is denoised by the noise variance $\sigdeux{$\eta$}(i,p)$ estimated by interpolating the auto-covariance peak. Finally, we evaluate $\tau$ as the FWHM of $\overline{\gamma}(\Delta t)$. The wind-speed depends on the ratio $d/\tau$ where $d$ is the sub-aperture size in the pupil plane, actually 0.6~m for \cana in phase B. Our purpose is now to calibrate the relation between the real wind-speed on-sky assuming a Taylor-like turbulence and the ratio $d/\tau$. 

Using the end-to-end simulation tool YAO~(https://github.com/frigaut/yao.git), we have generated phase screens with nominal seeing~($r_0 = 20$ cm) and outer scale~($L_0 = 9$ m) with different wind-speed values and wind direction comprised between 0 and 2$\pi$. We use our routine to determine the ratio $d/\tau$ and compare this value to the set wind-speed in YAO to get the factor to be applied to $d/\tau$.

\begin{figure}
	\begin{center}
		\includegraphics[scale=.75]{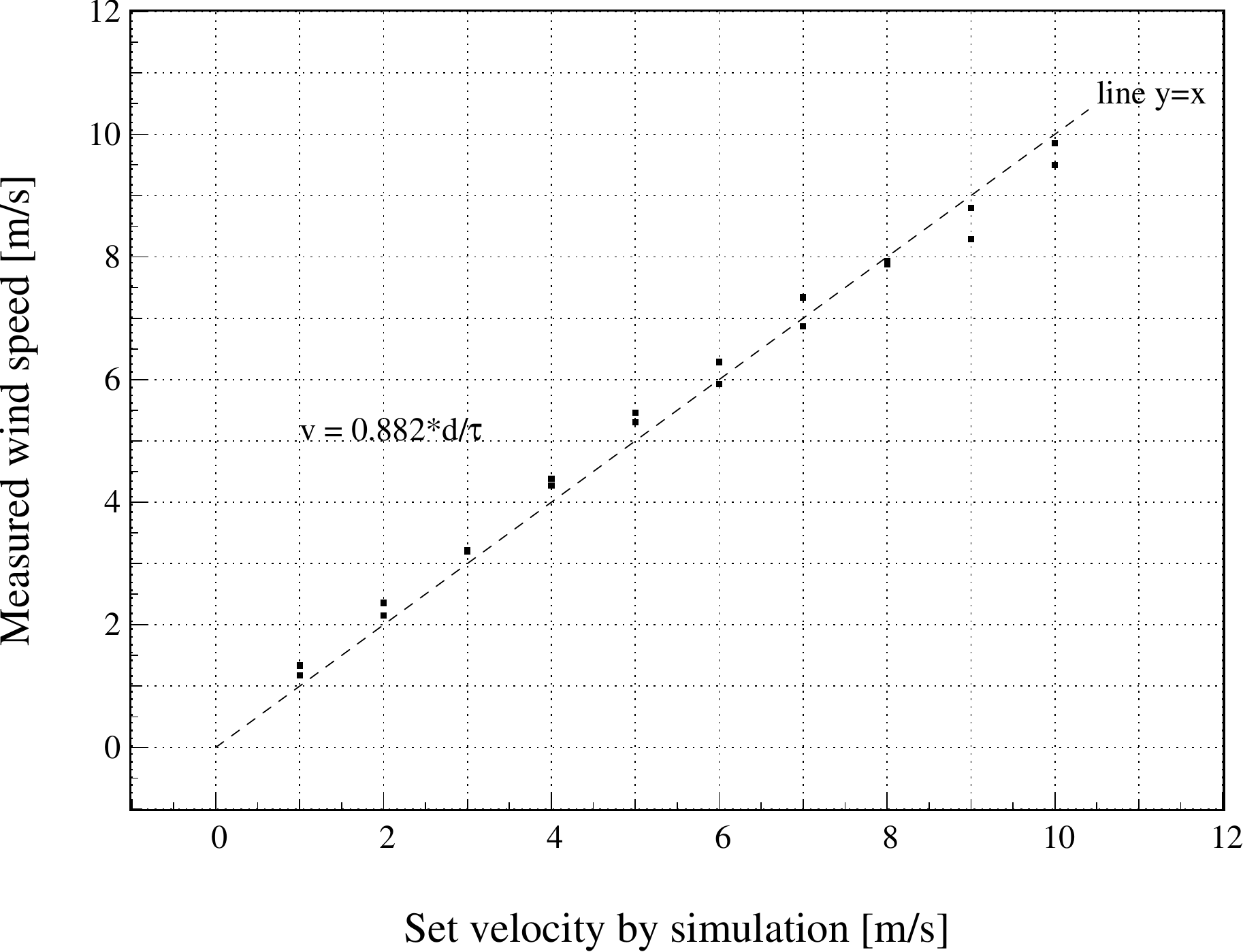}
		\caption{\small{Estimated wind velocity using Eq.~\ref{E:vcal} as function of YAO values. Each group of samples located at a single simulation velocity corresponds to a different phase screen with a different wind direction.}}
		\label{F:velcal}
	\end{center}
\end{figure}

In order to get as close as possible to the on-sky conditions, we have created phase screens of size 50~m times 2048~frames(our acquisition time on-sky). We have performed this comparison for wind-speed spanning from 0 to 10 m.s$^{-1}$, which corresponds to the range of values we have to deal with on-sky using \cana. At each velocity value, YAO has generated eight different phase screens from which we acquire system telemetry without noise. We expect to include the convergence effect of measurements into our calibration. Using a proportional regression of the ratio $d/\tau$ versus $v$, we get the following calibration:
\begin{equation} \label{E:vcal}
\widehat{v} \simeq 0.882 \times \dfrac{d}{\tau}.
\end{equation}
We report in Fig.~\ref{F:velcal} the wind-speed estimated using Eq.~\ref{E:vcal} versus the set value in YAO. We consider this calibration as acceptable for our next PSF-R purpose.

\subsubsection{MMSE layer-by-layer reconstruction}
\label{SS:MMSE}

To get the wind-speed profile, we perform a MMSE reconstruction of WFS slopes at a given layer. We need thus two covariance matrices to derive the reconstructor $\R(\hl)$ to get the phase at an altitude $h$~:
\begin{equation} \label{E:Rh}
\R(\hl) = \Sigma_{ss}(\hl) \times \para{\sum_{l=1}^{\nl}\Sigma_{ss}(\hl)}^{-1},
\end{equation}
where $\Sigma_{ss}(\hl)$ is the covariance matrix of all slopes, coming from both NGS/LGS WFS, for a sole turbulent layer at the altitude $\hl$. It is computed analytically using the model presented in Sect.~\ref{SS:model}. We then estimate the $\xth{l}$ turbulent layer contribution to the WFS measurements. 
\begin{equation}\label{E:Sh}
\widehat{\s}(\hl) = \R(\hl)\times \s
\end{equation}
Using the wind-speed estimation routine in Sect.~\ref{SS:FWHM} on $\widehat{\s}(\hl)$, we estimate the wind-speed at altitude $\hl$. 

\section{WHT site characterization} 
\label{S:stats}

With the L3S method (see Sect.~\ref{S1}) and our wind-speed profile retrieval~(see. Sect.~\ref{S:windspeed}), we have processed about 2,000 data sets acquired by \cana in May 23, 24 and 25, in July 17,18,19, 22, 23 and 24 and in September 13, 15, 16, 17 and 18 of September 2013. See Sect.~\ref{SS33} and~(\cite{Morris2014}) for more information about the \cana design. 

\begin{figure} 
	\begin{center}
		\includegraphics[scale=.45]{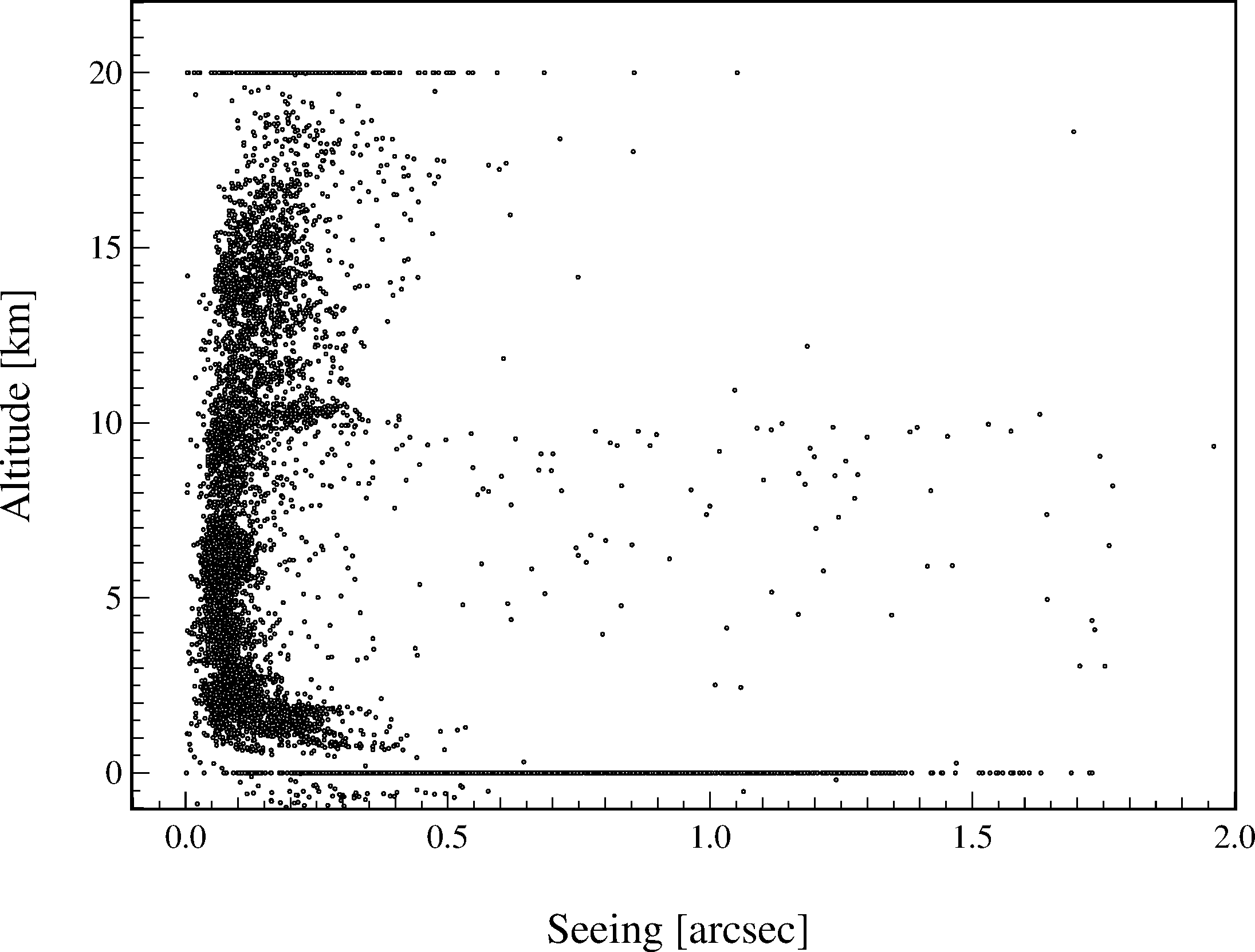}
		\hspace{.15cm}
		\includegraphics[scale=.45]{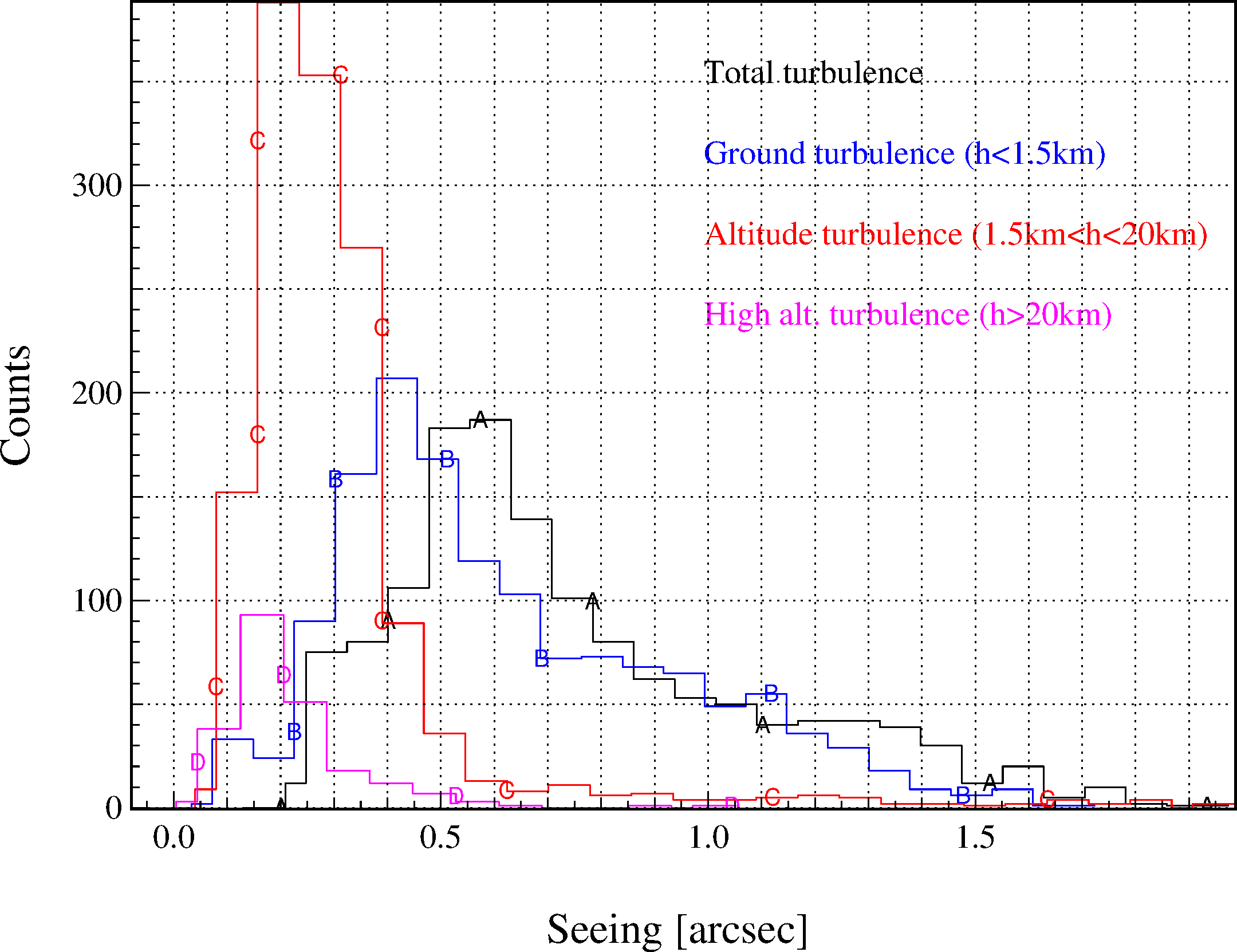}
		\vspace{.75cm}
		
		\includegraphics[scale=.45]{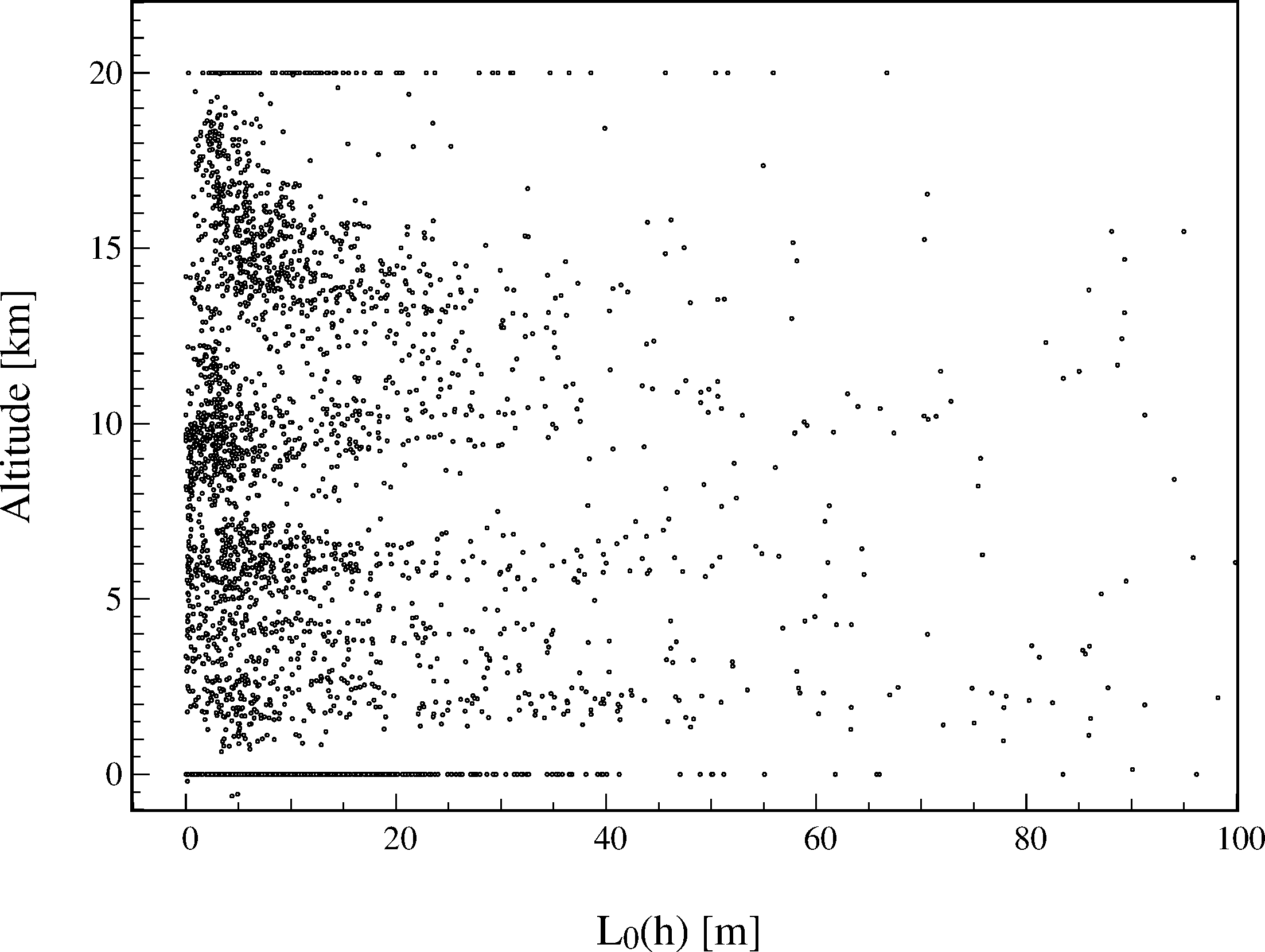}
		\hspace{.15cm}
		\includegraphics[scale=.45]{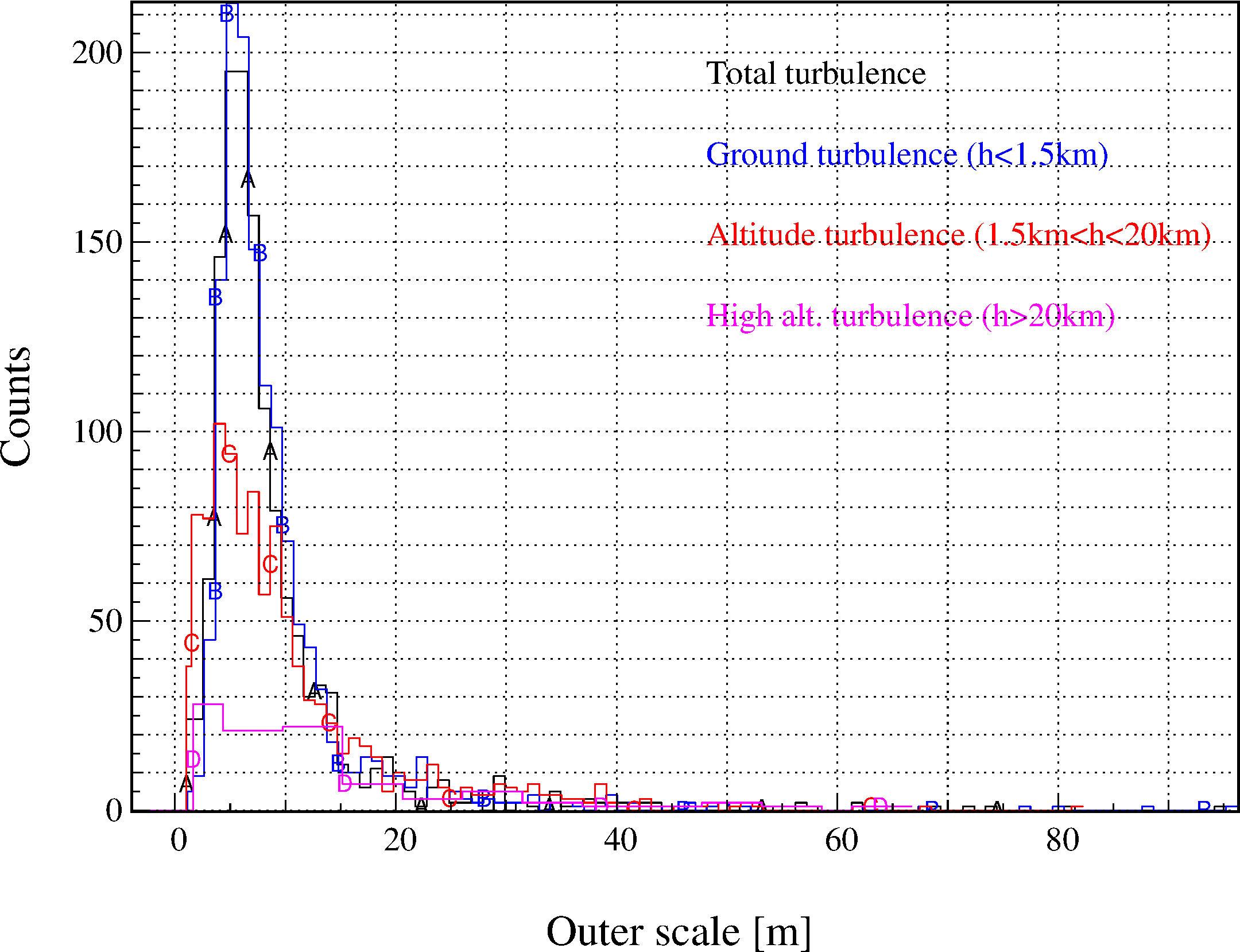}
		\vspace{.75cm}
			
		\includegraphics[scale=.45]{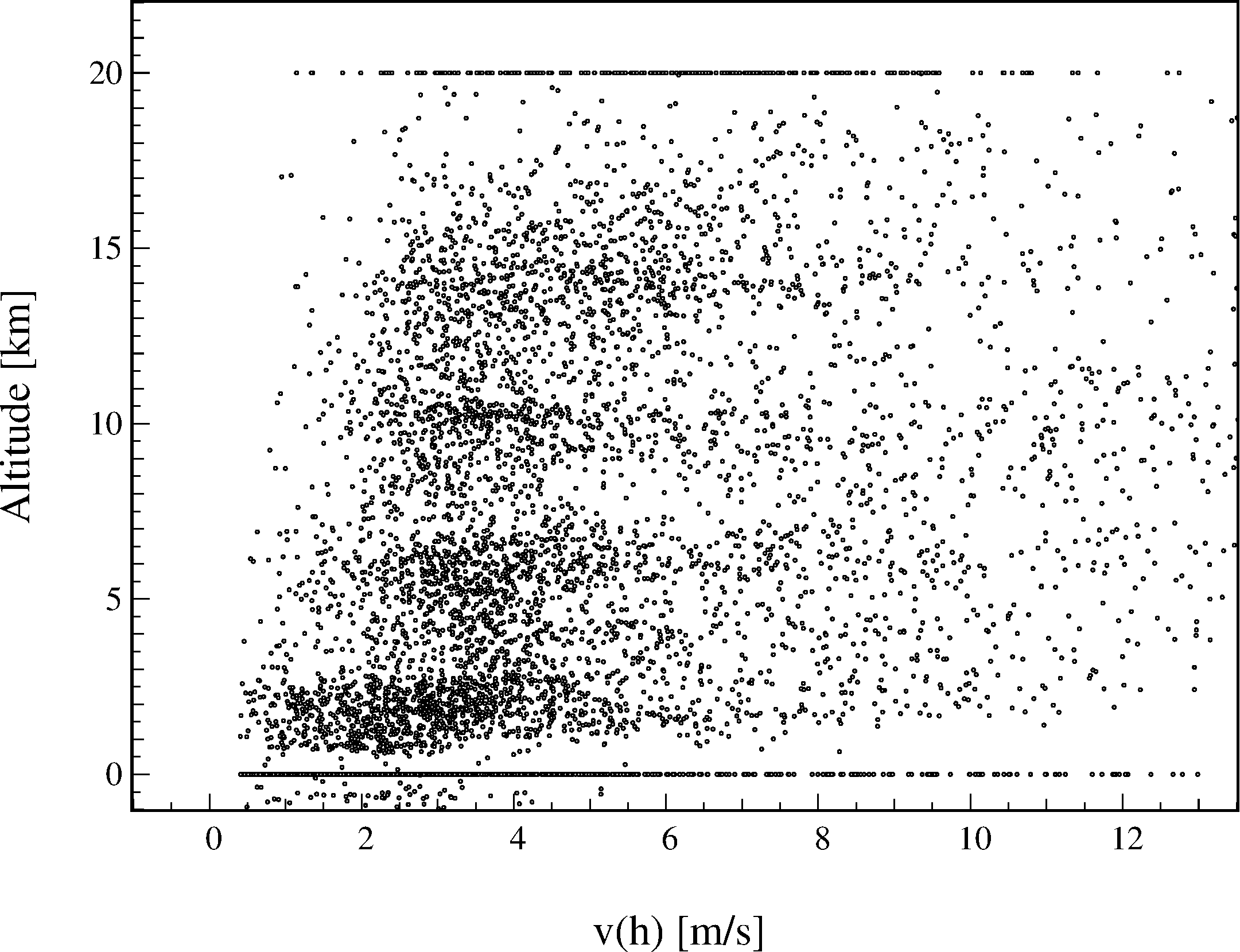}
		\hspace{.15cm}
		\includegraphics[scale=.45]{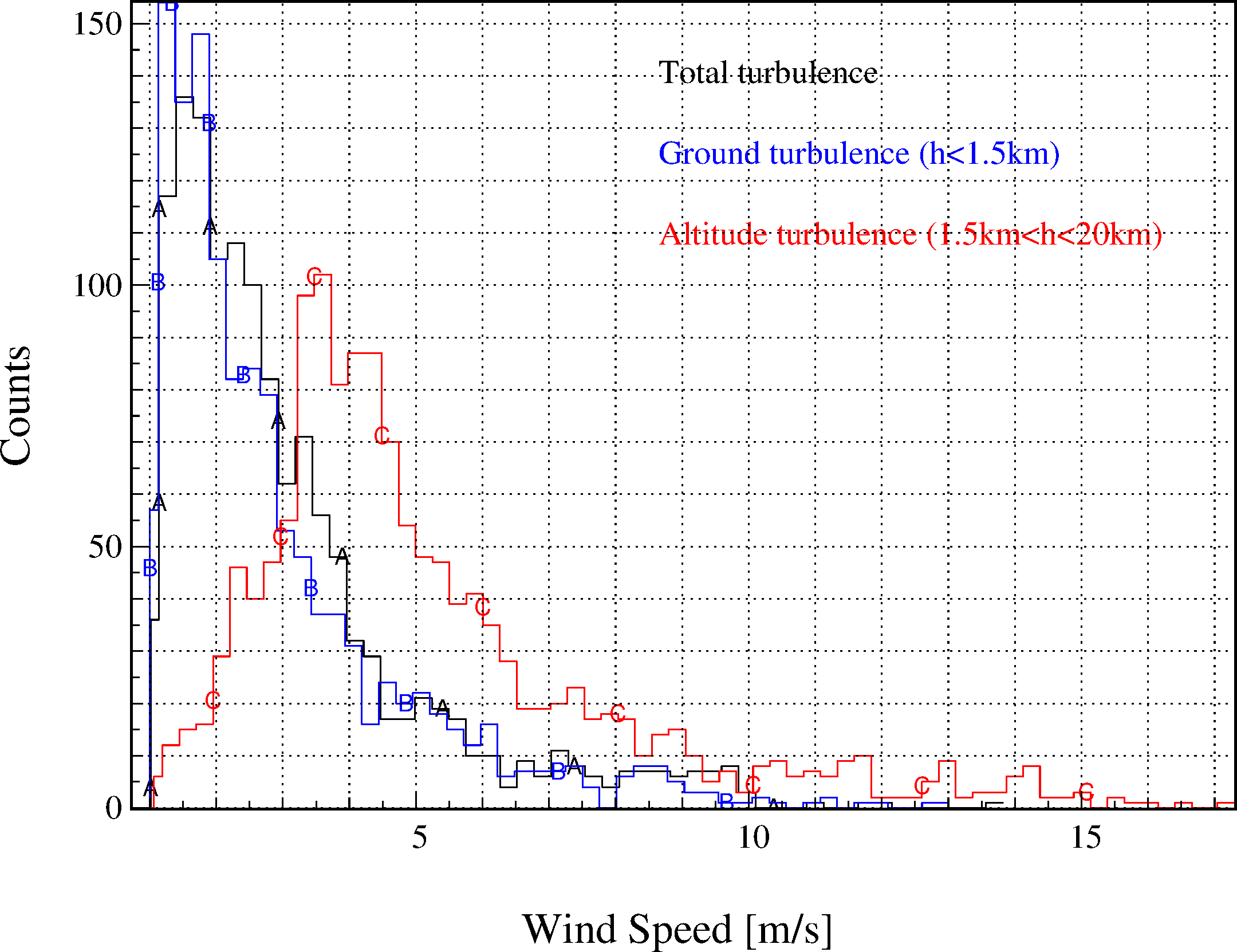}
		
		\caption{\small{\textbf{Left:} Distribution of seeing, outer scale and wind-speed in altitude. Each point is a retrieved layer from one of the 5-layers profiles determined on 2,000 \cana data sets. \textbf{Right:} Histogram of total, ground~(below 1~km) and altitude turbulence characteristics. Each value has been averaged~(see Eq.~\ref{E:l0eff} and~\ref{E:veff} ) over either ground turbulence~(below 1~km) or altitude turbulence on a single data sets. The procedure has been repeated for all the 2,000 processed data sets. Values of seeing are unbiased from the airmass, as well as the altitude, and given at 500~nm of wavelength. All altitude layers retrieved above 20~km have been gathered into the plateau at 20~km. }}
		\label{F:cnhprof}
	\end{center}	
\end{figure}

On each data set, we have retrieved the seeing, outer scale and wind-speed profiles over five layers. The left panels of Fig.~\ref{F:cnhprof} show the distribution of the turbulence along the altitude as a kind of a probability density. Each point is a retrieved layer. 

Using the auto-correlation of slopes, it is possible to get an estimation of the seeing above 20~km, the maximal altitude we can reconstruct by tomography with \textsc{Canary}. We have gathered the seeing of such very high altitude layers into the 20~km plateau in Fig.~\ref{F:cnhprof}.

In addition, we report on the right panels of Fig.~\ref{F:cnhprof} histograms of seeing, outer scale and wind-speed. We have split the turbulence into a ground part, altitude below 1~km, and an altitude part. For each retrieved turbulent profile, we have then averaged the ground and altitude values to analyze the turbulence behavior, differently if it comes from the ground or the altitude. To do that, we define the \emph{effective outer scale} as~:
\begin{equation} \label{E:l0eff}
L_0^* = \para{\dfrac{\sum_l L_0^{-5/3}(\hl) \times \rz^{-5/3}(\hl)  }{\sum_l \rz^{-5/3}(\hl)}}^{-3/5},
\end{equation}
and the \emph{coherent wind-speed}, coming from the coherent time of the turbulence, as:
\begin{equation} \label{E:veff}
v_0^* = \para{\dfrac{\sum_l v^{-5/3}(\hl) \times \rz^{-5/3}(\hl)  }{\sum_l \rz^{-5/3}(\hl)}}^{-3/5},
\end{equation}
where the summation is done over all layers or altitude layers~(above 1~km) or ground layers only. Eq.~\ref{E:l0eff} and Eq.~\ref{E:veff} gives integrated values of outer scale and wind speed in averaging the profile along layers in a same way than the $r_0$ computation~(\cite{Roddier1981}).

The $\rz^{-5/3}$ is retrieved for any layers using the L3S approach. We report in Tab.~\ref{T:statturbu} average values for seeing, layers contribution, effective outer scale and coherent wind-speed. 

\begin{table}[h!]
	\centering
	\begin{tabular}{|c|c|c|}
		\hline
		& Median values  & 1-$\sigma$ std values\rule[-7pt]{0pt}{20pt}	 \\
		\hline
		Global seeing & 0.699~"  & 0.35~"\rule[-2pt]{0pt}{12pt}	\\
		Ground seeing (h$\leq$1.5~km)& 0.552~" &0.32~"  \rule[-2pt]{0pt}{12pt}\\
		Altitude seeing (1.5~km$>$h$\leq$20~km) & 0.263~" &0.23~" \rule[-2pt]{0pt}{12pt}\\
		High Altitude seeing (h$>$20~km, appears in 16\% of processed cases) & 0.187~" & 0.13~" \rule[-2pt]{0pt}{12pt}\\
		\hline
		Ground contribution & 76.8\% & 17\%\rule[-2pt]{0pt}{12pt}\\
		Altitude contribution & 22.2\% & 16.4\%\rule[-2pt]{0pt}{12pt}\\
		High altitude contribution & 7.46\% & 14.9\% \rule[-2pt]{0pt}{12pt}\\
		\hline
		Global $L_0^*$ & 8.9~m & 8.4~m\rule[-2pt]{0pt}{12pt}\\
		Ground $L_0^*$ & 9.2~m  & 7.9~m\rule[-2pt]{0pt}{12pt}\\
		Altitude $L_0^*$ & 10.3~m &  9.7~m\rule[-2pt]{0pt}{12pt}\\
		High Altitude $L_0^*$ & 12.9~m &  13.1~m\rule[-2pt]{0pt}{12pt}\\
		\hline
		Global $v_0^*$ & 3.07~m.s$^{-1}$ &  1.9~m.s$^{-1}$\rule[-2pt]{0pt}{12pt}\\
		Ground $v_0^*$ & 2.89~m.s$^{-1}$  & 1.9~m.s$^{-1}$\rule[-2pt]{0pt}{12pt}\\
		Altitude $v_0^*$ & 5.22~m.s$^{-1}$ &  2.78~m.s$^{-1}$\rule[-2pt]{0pt}{12pt}\\
		\hline
	\end{tabular}
	\caption{ \small{ Statistics on seeing at 500~nm, effective outer scale~(see Eq.~\ref{E:l0eff}) and coherent wind-speed~(see Eq.~\ref{E:veff}). It gives total values averaged over 2,000 data sets with ground and altitude layers contribution. Altitude and seeing are unbiased from the telescope airmass. Statistics on outer scale are given in throwing all outer values larger than 100~m, the software breakdown above which the outer scale retrieving is stuck.}}
	\label{T:statturbu}
\end{table}

As expected, the ground layer is the most dominant one with a median value of 76.8\% of the total turbulence, while altitude layers, below 20 km represent about 22\% of the turbulence. We get on average a ground seeing of 0.67", which is close to the reported value at the WHT~(\url{http://www.ing.iac.es/Astronomy/telescopes/wht/}).

According to histograms in Fig.~\ref{F:cnhprof}, the ground layer also dominates the seeing variability. Altitude layers below 20~km reach an average seeing of 0.26" and stay relatively constant in strength with time. We also notice negative altitude layers. They may appear between the WHT primary and secondary mirror, but we cannot determine if they really exist or if this is a LMA artifact.

Note the very altitude layers, above 20~km and too high for being reconstructed by tomography with \cana, reach 7.5\% of the total turbulence in median value, with 0.187" of seeing. However this only occurs in 16\% of cases. It means this phenomena is not so frequent, but significant in terms of seeing. \\

Fig.~\ref{F:cnhprof} displays the vertical outer scale distribution. The relevance of the outer scale for tomography and PSF reconstruction is one of the actual preoccupation~(See. paper 9909-16 \textit{Review on AO real-time turbulence estimation} in that conference).  The plot is less dense than the ones from the seeing and wind-speed distribution since 55\% of samples have reached more than 100~m of outer scale. This is a software breakdown in the L3S code: above 100~m of outer scale, we no longer consider these is any physical meaning. 

From the outer scale distribution, we draw two other observations: firstly, in more than 55\% the L3S~(and generally the L\&A approach) retrieves an outer scale larger than 100~m. This is a software breakdown we have implemented into the spatial covariance model~: if the LMA reaches this limit, we stick the value to 100~m, we consider there is no more physical meaning on that.
In excluding those 55\% of cases, we have 6\% of cases that reach more than 20~m, 2.7\% with more than 30~m and only 1\% of cases with an outer scale greater than 50~m. 

The question now is why in 55\% of cases is the L\&A approach retrieved a fake value of the outer scale ? 

A main reason is we are observing with a 4.2~m telescope and we are not sensitive to large values of outer scale. But now, we can wonder whether we are reaching the software breakdown of 100~m because the aperture-size of the telescope or because any other reason to be determined using end-to-end simulation.\\

Secondly, this distribution of $L_0(h)$ is quite constant around twice the telescope diameter and is constrained by the ground layer. Tab.~\ref{T:statturbu} shows $\lz^*$ becomes larger in altitude compared to the ground value. The conclusion on that is the estimation of the outer scale we have is biased by the finite aperture size from which we observe the turbulence.

In altitude, because the stars angular separations, the meta-pupil size is increasing with the altitude. It is exactly like the aperture-size through which we observe the turbulence is increasing as well, as the outer scale in altitude looks to do according to Fig.~\ref{F:cnhprof}. 

From those two observations, we conclude the estimation of the outer scale we are doing on a 4.2~m telescope is not directly related to a physical quantity. The problem we have now is to determine whether taking into account the entire outer scale profile is relevant for tomography and PSF-reconstruction on \cana and demonstrate and understand why, using end-to-end simulations, the outer scale, as we have defined in Eq.~\ref{E:l0eff}, is twice the meta-pupil size.

Figure~\ref{F:cnhprof} illustrates the wind-speed profile as well. It appears the ground turbulence is the slowest one, with 2.89 m.s$^{-1}$, against 5.22 m.s$^{-1}$ in altitude. We get a very strong variability of wind-speed values, especially in altitude, it makes thus the L3S approach more valuable.\\

Finally, the distributions of seeing, outer scale and wind-speed in Fig.~\ref{F:cnhprof} are denser around particular altitudes. These are the most contributing layers to the turbulence and are located at the ground, around 2~km, 6~km, 10~km and 16~km. The tomographic resolution we have on \cana allows to retrieve on average five turbulent layers.

\section{Conclusions}
\label{S:conclusions}

We have discussed on the L3S approach, an upgrade of the Learn\& Apply algorithm, which dissociates the identification of the altitude layers from the ground layer. Under nominal observation conditions, the parameters it estimates lead to the same level of tomographic error than the L\&A approach in using five time less temporal frames than the latter. Moreover, we can gain up to 30\% on the tomographic error using the L3S.

The L3S technique has been applied over a large set of \cana data to characterize the turbulence above the William Herschel Telescope~(WHT): we find 0.67"/8.9m/3.07m.s$^{-1}$ of total seeing/outer scale/wind-speed against 0.55"/9.2m/2.89m.s$^{-1}$ below 1.5~km and 0.263"/10.3m/5.22m.s$^{-1}$ between 1.5 and 20 km. We have also determined that the high altitude layers above 20~km, that cannot be compensated by \textsc{Canary}, appear only 16\% of time with a median seeing of 0.187". We have observed the outer scale values we can measure with \cana are limited by around twice the meta-pupil size at a given altitude.

\acknowledgments 
The research leading to these results received the support of the
A*MIDEX project (no. ANR-11-IDEX-0001-02) funded by the ”Investissements
d'Avenir” French Government program, managed by the French National
Research Agency (ANR). This work is supported by CNRS, INSU, Observatoire de Paris, Universit\'e Paris Diderot-Paris 7 and the European Commission (Fp7 Infrastructures 2012-1, OPTICON Grant 312430, WP1).

\end{document}